# Solvation of the Fluorine Containing Anions and Their Lithium Salts in Propylene Carbonate and Dimethoxyethane

Vitaly Chaban[1]

1) Instituto de Ciência e Tecnologia, Universidade Federal de São Paulo, 12231-280, São José dos Campos, SP, Brazil

2) Department of Chemistry, University of Southern California, Los Angeles, CA 90089, United States

**Abstract**. Electrolyte solutions based on the propylene carbonate (PC)-dimethoxyethane (DME) mixtures are of significant importance and urgency due to emergence of lithium-ion batteries. Solvation and coordination of the lithium cation in these systems have been recently attended in detail. However, analogous information concerning anions (tetrafluoroborate, hexafluorophosphate) is still missed. This work reports PM7-MD simulations (electronic-structure level of description) to include finite-temperature effects on the anion solvation regularities in the PC-DME mixture. The reported result evidences that the anions appear weakly solvated. This observation is linked to the absence of suitable coordination sites in the solvent molecules. In the concentrated electrolyte solutions, both $BF_4^-$ and $PF_6^-$ prefer to exist as neutral ion pairs ($LiBF_4$, $LiPF_6$).

---

[1] E-mail: vvchaban@gmail.com

**Introduction**

Electrolyte solutions occupy an honorable place in modern chemistry and material science due to their importance for multiple energy storage devices.[1,2,3,4-9] The energy storage devices, in turn, gain urgency thanks to emergence of portable devices, which all require reliable, safe, and durable power supplies. Lithium salts[1,10-12] dissolved in the mixed organic solvents, such as dimethoxyethane (DME) and propylene carbonate (PC) are applied in the lithium-ion batteries,[13-15] since they exhibit wide electrochemical windows and other favorable properties. High ionic conductivity and dielectric constant of the electrolyte solution always arrive at the cost of a high shear viscosity.[15] The compromise can be achieved by mixing highly polar cyclic ethers, such as PC, with low-polar linear ethers, such as DME. Fortunately, these solvents are well miscible with one another giving rise to a variety of compositions. The resulting liquid mixtures provide an electrochemically interesting background for electrolyte systems.

The difficulties of data interpretations in the case of many-atom molecules considerably hinder development of the efficient electrolyte solutions. The factors determining key properties of the electrolyte, such as ion solvation regularities and ion transport regularities, must be clearly understood to foster progress in the lithium-ion batteries. Mixed solvents are challenging both for theoretical and experimental investigations because of the abundant specific molecular interactions in these systems. The complexity increases drastically with every new molecular or ionic species added to the mixture.

The structure and dynamics of lithium ion in the PC-DME mixtures have been recently addressed in details.[2,3,12,14,16] The dependences on the salt concentration and temperature have been investigated. Although the agreement between different computational approaches (for instance, empirical-potential molecular dynamics and density functional theory based molecular dynamics)[2,14] does not seem to be achieved, the existing results foster discussion and understanding of these complicated ion-molecular systems. In turn, the structure and solvation of anions, such as $BF_4^-$, $PF_6^-$, still remain beyond the scope of researchers' interest. It follows from conventional chemical wisdom that anions are solvated more weakly, as compared to the lithium cation. It is, however, unclear how weakly. Information about solvation of anion is as important as that about cation, since solvation is responsible for proper dispersion of ions throughout the electrolyte solution, which ultimately determines the magnitude of ionic conductivity.

This work reports PM7-MD simulations on a set of the negatively charged and neutral non-periodic systems to investigate (1) structure and composition of the solvation shells surrounding tetrafluoroborate, $BF_4^-$, and hexafluorophosphate, $PF_6^-$, anions; (2) effect of the lithium-ion on the coordination number of the $BF_4^-$ and $PF_6^-$ anions; (3) energetics of the $BF_4^-$ and $PF_6^-$ anions binding to both co-solvent molecules. The PM7-MD method[17-20] provides a comprehensive tool offering an electronic-structure level of description of every system. Thanks to intelligently parameterized integrals in PM7,[21-24] PM7-MD performs more efficiently than density functional theory powered molecular dynamics simulations in terms of computational cost.

**Methodology**

The PM7-MD simulations[17-20] were performed using the eight systems, as represented in Table 1. Systems I-IV contain a single anion ($BF_4^-$ or $PF_6^-$) and a single solvent molecule (DME or PC). Such simulation setups are convenient to investigate anion-solvent interactions and compare them to one another. Systems V-VI contain a single anion ($BF_4^-$ or $PF_6^-$) in the equimolar PC-DME mixture. These systems allow representing an infinitely diluted solution of the corresponding anion. That is, the cation effect can be removed. In turn, systems VII-VIII contain an ion pair ($LiBF_4$ or $LiPF_6$) immersed in the same PC-DME mixture. These two systems account for the cation effect in the solvation of the anion.

Prior to PM7-MD simulations at 300 K, the geometries of systems I-IV were optimized using the improved eigenfollowing (EF) geometry optimization algorithm.[24] All PM7-MD simulations were performed at 300 K with temperature maintained constant using weak coupling to the external thermal bath.[25]

Table 1. Simulated systems, their fundamental properties and selected simulation details. Each system was started with three different configurations (arbitrarily arranged molecules) to ensure that the equilibrated ion-molecular configuration does not depend on the starting configuration. Proper equilibration of all systems was thoroughly controlled by analyzing evolution of many thermodynamic quantities, such as energy, dipole moments of molecules, selected non-covalent distances. Note that we provide the number of explicitly treated electrons rather than the total number of electrons in the corresponding MD system. PM7 represents valence electrons explicitly, while core electrons are considered using effective potentials

| # | Composition | # atoms | # explicit electrons | Type of calculation | Equilibration time, ps | Sampling time, ps |
|---|---|---|---|---|---|---|
| I | $1BF_4^-$ +1DME | 21 | 70 | OPT+MD | 2.0 | 8.0 |
| II | $1BF_4^-$ +1PC | 18 | 72 | OPT+MD | 2.0 | 8.0 |
| III | $1PF_6^-$ +1DME | 23 | 86 | OPT+MD | 2.0 | 8.0 |
| IV | $1PF_6^-$ +1PC | 20 | 88 | OPT+MD | 2.0 | 8.0 |
| V | $1BF_4^-$ +6DME+6PC | 179 | 500 | MD | 15 | 75 |
| VI | $1PF_6^-$ +6DME+6PC | 181 | 516 | MD | 15 | 75 |
| VII | $1LiBF_4$+6DME+6PC | 180 | 500 | MD | 15 | 75 |
| VIII | $1LiPF_6$+6DME+6PC | 182 | 516 | MD | 15 | 75 |

The PM7-MD method obtains forces acting on every atomic nucleus from the electronic structure computation using the PM7 semiempirical Hamiltonian.[21-24] PM7 is a parameterized Hartree-Fock method, where certain integrals are pre-determined based on the well-known experimental data, such as ionization energies. This solution allows for effective incorporation of the electron-correlation effects, while preserving a quantum-chemical nature of the method. Therefore, PM7 is able to capture any specific chemical interaction. On the contrary, classical pairwise interaction potentials are unable to represent formation/destruction of the covalent bonds, whereas formation/destruction of the hydrogen bonds can be modeled using point electrostatic charges. PM7 is more physically realistic than any existing force field based technique. Note that PM7 includes an empirical correction for the dispersive attraction. Thus, it can be successfully used to model hydrocarbon moieties. The accuracy and robustness of the

PM7 semiempirical parameterization, as applied to thousands of versatile chemical systems, was demonstrated by Stewart elsewhere.[21-24]

The derived forces are coupled with the initial positions of atoms and randomly generated velocities (Maxwell-Boltzmann distribution). Subsequently, Newtonian equations-of-motions can be constructed and numerically integrated employing one of the available algorithms. This work relies on the velocity Verlet integration algorithm. This integrator provides a decent numerical stability, time- reversibility, and preservation of the symplectic form on phase space. Due to rounding errors and other numerical inaccuracies, total energy of the system is not perfectly conserved, as in any other MD simulation method. Temperature may need to be adjusted periodically by rescaling atomic velocity aiming to obtain the required value of kinetic energy with respect to the number of degrees of freedom. This work employs a weak temperature coupling scheme with a relaxation time of 50 fs, whereas the integration time-step equals to 0.5 fs.[25] The chosen integration time-step is set based on the preliminary benchmarks on proper energy conservation (using molecular dynamics simulations in the constant-energy ensemble). Unlike in many applications of classical molecular dynamics, we do not constrain C-H covalent bond lengths (and corresponding valence angles). Therefore, an integration time-step must be set smaller to properly account for these vibrations.

More details of the present PM7-MD implementation are provided elsewhere.[17-20] The method has been successfully applied to address problems of ionic liquids[17,18] and nanoparticles.[19,20] Local structure of the liquid-matter ion-molecular systems was characterized using a set of pair correlation functions (PCFs). PCF shows how much number density of given atom at given separation from the reference atom is larger than an average number density of this atom in the considered system. Since the simulated PM7-MD systems are non-periodic, normalization of PCFs is, in principle, ambiguous. In this work, the volume of the simulated systems was formally admitted to be unity to perform normalization (the normalization is important to compare PCFs to one another). The PCFs were calculated using simple in-home tools along the sampling stage of each PM7-MD trajectory, 75 ps long (Table 1).

**Results and Discussion**

The optimized geometries of the simulated chemical objects (DME, PC, $BF_4^-$, $PF_6^-$) are depicted in Figure 1. PC and DME molecules are similar in terms of total number of electrons and atomistic composition. However, the structure formulas are different justifying their different physical chemical and thermodynamics properties.

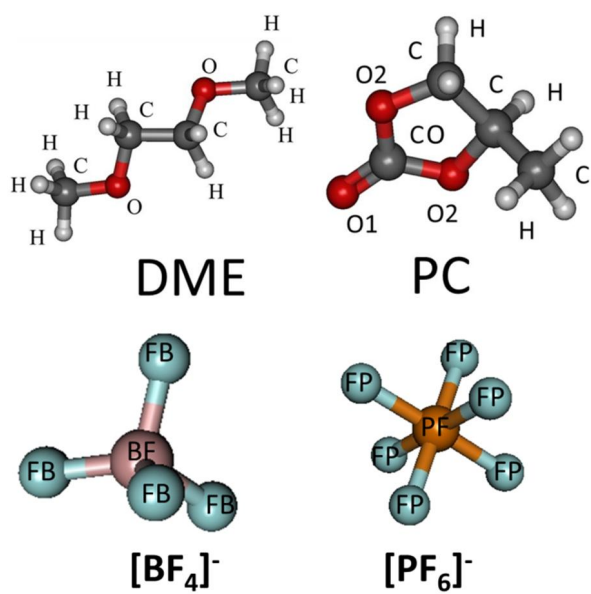

Figure 1. Optimized chemical structures of dimethoxyethane, propylene carbonate molecules, tetrafluoroborate ($BF_4^-$) and hexafluorophosphate ($PF_6^-$) anions. The geometry optimizations were carried out using the EF algorithm.[24] The wave function was constructed using the PM7 Hamiltonian approximation. These designations of atoms will be used throughout the paper to discuss the simulated systems.

The mixture of PC and DME is successful in the electrolyte solutions, because DME efficiently decreases shear viscosity of PC. This effect is only possible thanks to their good mutual miscibility, even though many physical chemical properties of these solvents differ significantly. According to the Coulson-type population analysis, symmetric oxygen atoms, O, in DME are moderately electron-rich, -0.39e each. Carbon atoms are also slightly negative: -0.06e in the ethane backbone and -0.20e in the -O-CH$_3$, moieties. An entire molecule is neutralized by the ten hydrogen atoms, which are all positively charged, +0.12-0.15e. Interestingly, all three oxygen atoms of PC also carry -0.39e charges, exactly as in the case of DME. The corresponding electrostatic attraction is the major prerequisite of a good miscibility featured by PC and DME. The most strongly electrostatically charged atom in PC is the carbonate carbon, CO, +0.69e. The dipole moment of PC, evaluated from the electrostatic potential, equals to 4.6 D, which is in quite a decent agreement with the experimental value, 4.9 D. Note that the experimental dipole moment must be higher due to an electronic polarization in the condensed phase, whereas the present calculation considers just a single PC molecule in vacuum.

The geometries of the negatively charged anion-molecular complexes — [BF$_4$-DME]$^-$, [BF$_4$-PC]$^-$, [PF$_6$-DME]$^-$, and [PF$_6$-PC]$^-$ — require 200-350 EF cycles to reach the local minimum states on the corresponding potential energy surfaces. The optimizations were started from a few arbitrary initial configurations to locate, possibly, the deepest minima of the complexes. The formation energies depicted in Figure 2 correspond to the most energetically favorable final anion-molecular configurations. Noteworthy, both BF$_4^-$ and PF$_6^-$ get better solvated in PC than in DME. The difference equals to 192 kJ mol$^{-1}$ in the case of BF$_4^-$ and to ca. 189 kJ mol$^{-1}$ in the case of PF$_6^-$. The found differences are significant, since they well exceed kT at room and elevated temperatures. However, these energies do not account for the entropy factor. Furthermore, they do not account for the composition of the first and second solvation shells of these anions in the condensed matter systems. The reported energies provide expectations of what must be expected in the PC-DME mixtures, although the energetic factor cannot be considered alone to generate reliable conclusions regarding larger systems.

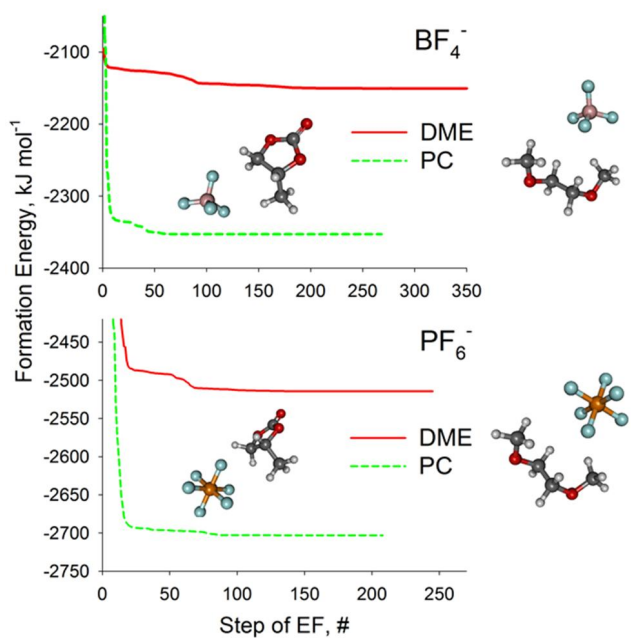

Figure 2. Evolution of formation energy upon geometry optimization of the [BF$_4$-DME]$^-$, [BF$_4$-PC]$^-$, [PF$_6$-DME]$^-$, and [PF$_6$-PC]$^-$ negatively charged complexes. The optimized geometries of these four complexes are shown as insets. The geometry optimizations were carried out using the EF algorithm. The wave function was constructed using the PM7 Hamiltonian approximation.

The optimized configurations of the anion-molecular complexes (Figure 2) suggest absence of coordination sites for the anions in both solvent molecules. While lithium cation is coordinated by the carbonate group of PC and the methoxy groups of DME,[14] BF$_4^-$ and PF$_6^-$ approach the hydrocarbon moieties of the solvents. These interactions are preferentially van der Waals in their nature. That is, anion-solvent binding must be expected to be weak.

The pathway to the closest local minimum requires smaller amount of the EF steps (cycles) ion the case of PC and PF$_6^-$, irrespective of other constituents of these systems (Figure 2). This observation is consistent with higher formation energies in the PC and PF$_6^-$ containing complexes. In addition, a large amount of the EF cycles required for the DME containing complexes is due to conformational flexibility of this molecule, as opposed to PC.

Figure 3 investigates fluctuations of the formation energy and compares the formation energies in the complexes of BF$_4^-$ and PF$_6^-$ with DME and PC separately (four independent complexes in total). All systems attain thermodynamic equilibrium within ca. 2 ps (Table 1). The performed analysis clearly indicates that the anions prefer to bind with PC. The lithium cation exhibits the same trend, as evidenced using classical and ab initio molecular dynamics simulations earlier.[2,14] The depicted data suggest either exclusive solvation of $BF_4^-$ and $PF_6^-$ by PC or at least preferential solvation by this solvent at finite temperature in the condensed phase. Larger-scale simulations are necessary to shed light on this important feature.

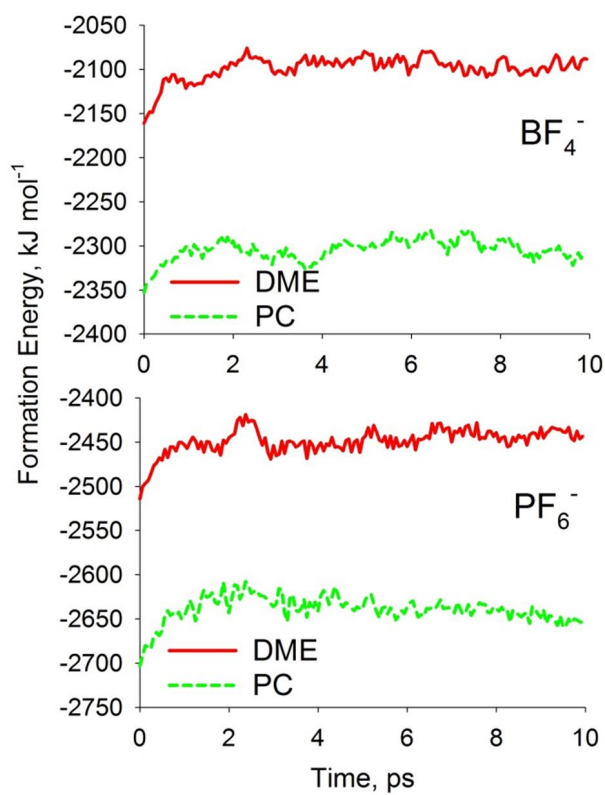

Figure 3. Fluctuations of formation energies of the negatively charged complexes. Top: $[BF_4\text{-DME}]^-$ (red solid line) and $[BF_4\text{-PC}]^-$ (green dashed line). Bottom: $[PF_6\text{-DME}]^-$ (red solid line) and $[PF_6\text{-PC}]^-$ (green dashed line). See systems I-IV in Table 1. The molecular trajectories were propagated during 10 ps with a time-step of 0.5 fs using the PM7-MD method. Before recording molecular trajectories, the formation energies of all negatively charged complexes were minimized using the EF algorithm. The increase of formation energies during the first picoseconds occurs due to heating from 0 to 300 K, because geometrically optimized structures correspond to zero-temperature.

Information about an ion behavior in a pure solvent is important to understand physical chemical and thermodynamics trends. However, such information is not sufficient to characterize solvation shells of an ion in the mixed solvent. Solvation is a complicated phenomenon, whereby not only enthalpy (potential energy factor) plays a role, but also entropic contribution to the solvation free energy is extremely important. In the case of sufficiently large molecules, entropic contribution can be even decisive. Based on Figure 3, one can hypothesize that the first solvation shell of $BF_4^-$ and $PF_6^-$ are populated by the PC molecules, while DME molecules are not located in the vicinity of the anions. In the meantime, we empirically know that DME and PC are miscible well finding routine application in the lithium-ion batteries. In the following, I will illustrate that the solvation shell composition can be only roughly predicted on the basis of enthalpy factor, whereas finite-temperature molecular dynamics simulations, such as PM7-MD, are absolutely necessary to observe a realistic distribution of molecules around the anions.

The PCFs derived between fluorine (FB, FP) atoms of $BF_4^-$ and $PF_6^-$ of carbon (C) atoms of PC and DME are provided in Figure 4. All functions indicate a weak solvation of the anions by PC and DME, despite the polarity of these solvents. Indeed, dielectric constant equals to 64 in PC and to 7.2 in DME. The dipole moment equals to 4.9 D in PC and to 1.7 D in DME. In the case of tetrafluoroborate anion, the first maximum for PC is located at 4.2 Å, whereas the first maximum for DME is located at 5.5 Å. In the case of hexafluorophosphate, both the PC and DME solvents feature peaks at ca. 4.8 Å.

Both solvent molecules are able to approach the anions as close as 2.1 Å. However, due to thermal motion, the solvent molecules do not stay at this position during most time. Clusters in vacuum cannot obviously represent a liquid matter system properly, although they provide an important understanding of the prospective structure regularities in those systems. Although the formation energies of the anion-PC clusters are somewhat higher, as compared to those of the anion-DME clusters, their effect on the pair correlation functions (Figure 4) is very modest.

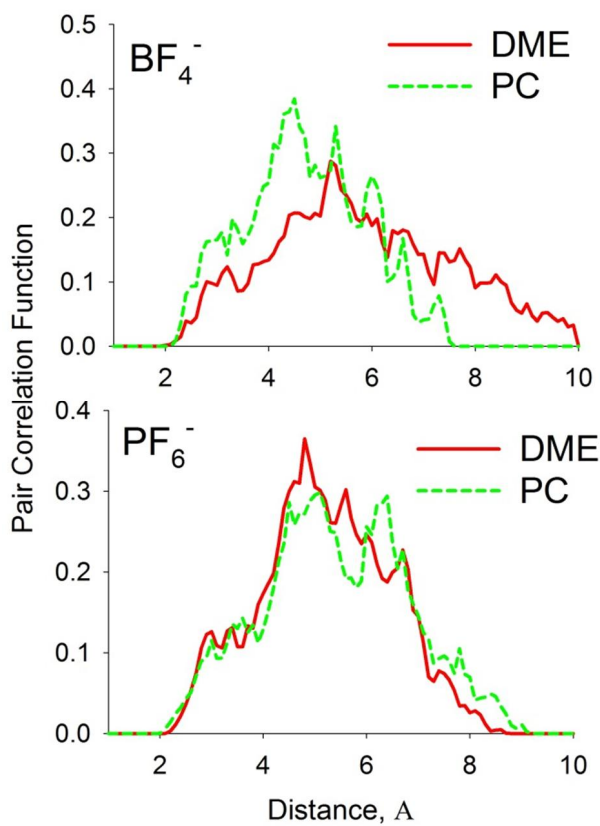

Figure 4. Pair correlation functions derived between fluorine (FB, FP) atoms of the $BF_4^-$ and $PF_6^-$ anions and carbon (C) atoms of PC and DME. See legends for designation. The depicted data are based on the PM7-MD simulations of systems I-IV. The functions were computed using the equilibrium parts of the trajectories (Table 1).

Figures 5-6 report PCFs in larger systems, V-VIII (Table 1). The PCFs were calculated between the fluorine atoms of the anions and carbon atoms of the solvent molecule, separately of PC and DME. The negatively charged (containing $BF_4^-$ and $PF_6^-$) and neutral (containing $LiBF_4$ and $LiPF_6$) systems are compared. In the case of the negatively charged systems, PC prevails in the first solvation shells of both fluorine containing anions. However, the anion appears in no case solvated exclusively by PC. Certain fraction of DME is always present nearby anions, which constitutes an important feature of these systems. Coordination number for each solvent separately can be deduced by integrating the first peak of the corresponding PCFs. Note that since the boundary of the first peak cannot be perfectly defined (meaning that the coordination sphere is diffuse), certain ambiguity persists in the reported coordination numbers. On the average, $BF_4^-$ features coordination number of 4.1 (PC) plus 1.5 (DME). $PF_6^-$ features coordination number of 3.7 (PC) plus 1.5 (DME). Recall that the investigated solvent mixture is equimolar. Therefore, the solvent composition in the vicinity of the anions alters significantly.

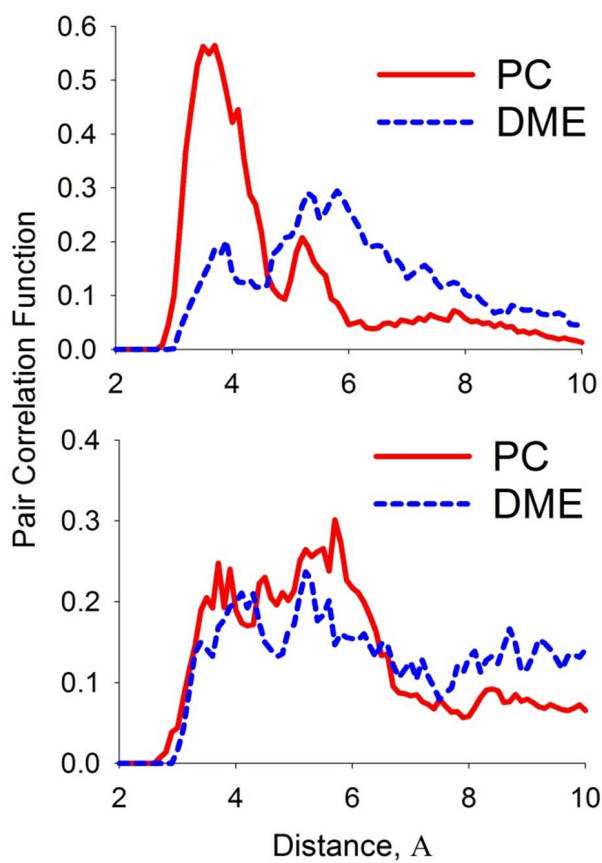

Figure 5. Pair correlation functions derived between the fluorine (FB) atoms of the $BF_4^-$ anion and carbon (C) atoms of PC and DME in the negatively charged ($BF_4$ + 6 DME + 6 PC]$^-$ (top) and neutral (LiBF$_4$ + 6 DME + 6 PC] (bottom) systems. See legends for designation. The depicted data are based on the PM7-MD simulations of systems V and VII. The functions were computed using the equilibrium parts of the trajectories (Table 1).

The lithium cation forms ion pairs with $BF_4^-$ and $PF_6^-$ at this salt concentration, which remain stable during the entire duration of the 75 ps long productive runs. The coordination number of $BF_4^-$, composing a neutral ion pair with the lithium cation, equals to 2.5 (PC) plus 1.7 (DME). These numbers can be directly compared to the case of $PF_6^-$: 2.8 (PC) plus 1.6 (DME). The cation removes approximately one solvent molecule from the coordination sphere of the fluorine containing anion. In all cases, this molecule appears propylene carbonate rather than dimethoxyethane. This result is in concordance with previous reports, which show that $Li^+$ pulls PC molecules towards itself. Overall, the $Li^+$ cation does not alter the existing solvation pattern of the anion.

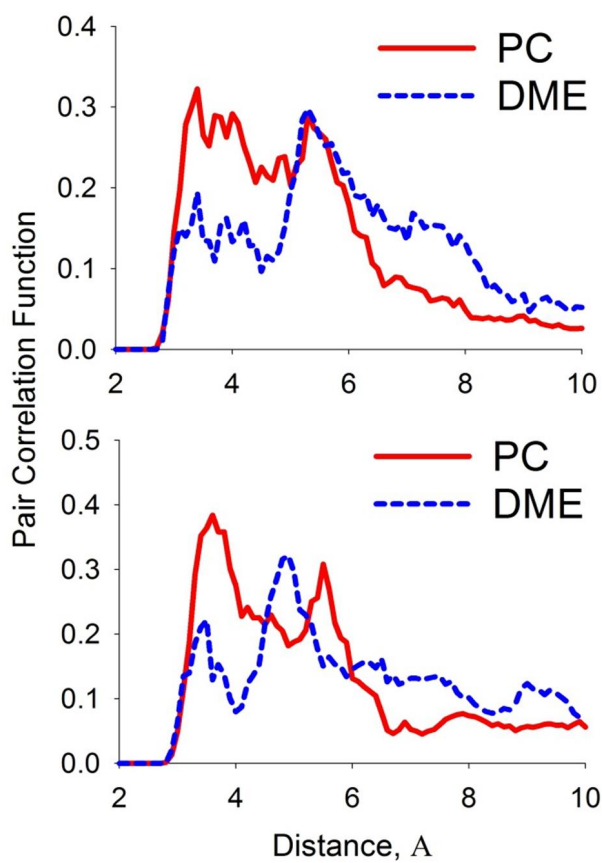

Figure 6. Pair correlation functions derived between the fluorine (FB) atoms of the $BF_4^-$ anion and carbon (C) atoms of PC and DME in the negatively charged $(PF_6 + 6\ DME + 6\ PC]^-$ (top) and neutral $(LiPF_6 + 6\ DME + 6\ PC]$ (bottom) systems. See legends for designation. The depicted data are based on the PM7-MD simulations of systems VI and VIII. The functions were computed using the equilibrium parts of the trajectories (Table 1).

To recapitulate, propylene carbonate is primarily responsible for solvation of both lithium cation and inorganic anions in the PC-DME mixtures. DME molecules also participate in the first solvation shell. However, both the energetics and the coordination numbers are inferior to those of PC. DME plays an important role is decreasing shear viscosity of the solution. The absence of suitable coordination sites for an anion in the solvent molecule and small dipole moment are the major responsible molecular-precision factors for that.

**Conclusions**

The paper reports PM7-MD simulations on a set of neutral and negatively charged systems containing propylene carbonate and dimethoxyethane molecules, tetrafluoroborate and hexafluorophosphate anions and lithium cation (to form an ion pair in solution with the corresponding anion). Since the solvation of lithium cation in the PC-DME mixtures was thoroughly attended before,[2,3,14] the present work concentrates on the solvation regularities of the $BF_4^-$ and $PF_6^-$ anions. These anions are most widely used in combination with $Li^+$ in the PC-DME mixtures to produce an efficient electrolyte system. Ion solvation phenomena are important for an application of these and similar systems as electrolytes in batteries. The PM7-MD method provides an electronic-structure level description including all phenomena, which are inaccessible for classical molecular dynamics. Certain electronic polarization may be expected in these anion-molecular systems/complexes, although, obviously, being weaker than in the case of lithium cation. Empirical Hamiltonians either ignore this or represent effectively following an average field approximation. The present PM7-MD results may be perceived as a benchmark for further larger-scale calculations.

The performed calculations suggest that both $BF_4^-$ and $PF_6^-$ exhibit generally weak binding to both PC and DME. Even the first solvation shells of both anions cannot be clearly determined. This solvation regularity is primarily due to absence of specific anion-coordination sites in the PC and DME solvent molecules. It is opposed to the clearly determined cation-coordination site, which is the carbonate group of PC. PC exhibits somewhat higher affinity to $BF_4^-$ and $PF_6^-$. The presence of the PC molecules in the poorly defined first solvation shells of these inorganic anions is systematically larger than the presence of the DME molecules.

The observations concerning the solvated anion structures are consistent with the thermodynamics analysis. However, conclusions derived from thermodynamic and structure analyses are not identical, since the former miss many-body and finite-temperature effects.

The results reported in the paper are important for understanding of modern lithium-ion batteries. They provide molecular principles for the optimization of the electrolyte properties, such as solvated ion structure and, further, ionic conductivity and shear viscosity, since the latter depend on the structure and thermodynamics of the solutions.

**Acknowledgments**

Prof. Vitaly Chaban is CAPES fellow under the "Science Without Borders" program.